\documentclass[12pt,preprint]{aastex}

\newcommand{\unit}[1]{\,\mathrm{#1}}
\newcommand{\kms}{\;{\rm km\,s^{-1}}}

\newcommand{\oMW}{{\scriptscriptstyle \rm MW}}

\newcommand{\sgn}{\mathrm{sgn}}

\newcommand{\fdeg}{\hbox{$.\mkern-4mu^\circ$}}

\newcommand{\vect}[1]{\mbox{\boldmath$#1$}}
\begin{document}
\title{The orbital poles of Milky Way satellite galaxies: a rotationally supported disc-of-satellites}

\author{Manuel Metz, Pavel Kroupa}
\affil{Argelander-Institut f\"ur Astronomie, Universit\"at
Bonn, Auf dem H\"ugel 71, D--53121 Bonn, Germany}
\email{mmetz,pavel@astro.uni-bonn.de}

\author{Noam I.\ Libeskind}
\affil{Racah Institute of Physics, Hebrew University of Jerusalem, Givat Ram, Jerusalem 91904, Israel}
\email{nil@phys.huji.ac.il}

\begin{abstract}
Available proper motion measurements of Milky Way (MW) satellite galaxies are used to calculate their orbital poles and projected uncertainties. These are compared to a set of recent cold dark-matter (CDM) simulations, tailored specifically to solve the MW satellite problem. We show that the CDM satellite orbital poles are fully consistent with being drawn from a random distribution, while the MW satellite orbital poles indicate that the disc-of-satellites of the Milky Way is rotationally supported. Furthermore, the bootstrapping analysis of the spatial distribution of theoretical CDM satellites also shows that they are consistent with being randomly drawn. The theoretical CDM satellite population thus shows a significantly different orbital and spatial distribution than the MW satellites, most probably indicating that the majority of the latter are of tidal origin rather than being DM dominated sub-structures. A statistic is presented that can be used to test a possible correlation of satellite galaxy orbits with their spatial distribution.
\end{abstract}


\section{Introduction}\label{sec_intro}
The Milky Way (MW) is surrounded by about 20 known satellite galaxies of which the very faint ones were found most recently by systematic search campaigns \citep{willm05,sakam06,zucke06c,zucke06,belok06,belok07,walsh07} in the Sloan Digital Sky Survey \citep[SDSS,][]{york00} catalogue data. These satellite galaxies are unique laboratories to study galaxy evolution since they are close enough to resolve individual stars, they are massive enough to have possibly kept back gas in the past, fuelling star formation processes, and they are old enough to study formation processes of galaxies in the early Universe and their evolution over a Hubble time. Yet, the true nature of the dwarf-spheroidal (dSph) satellites is still unclear.

The dSph satellite galaxies of the MW (and Andromeda) are mostly believed to be luminous, cold dark-matter (CDM) dominated subhaloes that fell into the Milky Way's host halo \citep{white78}. A lot of effort has recently been made to push the limits of CDM simulations to make predictions of luminous CDM sub-structures within MW sized host-haloes. Performing hierarchical structure-formation simulations combined with algorithms describing dark-matter -- luminous-matter biases, as given by semi-analytic galaxy formation models for example, DM sub-haloes are generated that appear to resemble dSph satellite galaxies \citep[e.g.][]{libes05,libes07,moore06}. Given the discrepancy between the observed visible matter and the inferred total mass, dSph satellites would be the most dark-matter dominated objects currently known in the Universe \citep{wilki02,kleyn03}. Their inferred dark matter profiles, however, are incompatible with CDM theory \citep{wilki06}. Recent observational constraints question the validity of the CDM model in these systems, and imply that perhaps the warm dark-matter (WDM) model may need to be considered as a viable alternative \citep{gilmo07}. This interpretation relies on solving the Jeans equation which implies that the dSph satellites have cores extending to about $100\unit{pc}$, whereas CDM haloes are known to be cuspy with a $\rho \propto r^{-p},\;p\approx 1$ -- 1.5, density profile \citep{diema05}. Independent evidence for the existence of cores come from observations of sub-structure within Ursa Minor \citep{kleyn03} and the placement of globular clusters in Fornax \citep{goerd06,sanch06}.

An alternative theory for the origin of dSph satellite galaxies of the MW is that they are of tidal origin: given the satellites fall on one or two great circles on the sky \citep{kunke76,lynde76}, \citet{lynde83} first suggested that the satellite galaxies of the MW might be torn off a common progenitor galaxy. With the advent of large N-body CDM simulations this idea was often overlooked, since in CDM simulations dSph galaxies are naturally identified with accreted cosmological sub-structures, even though not all of their properties could be fully understood (see above). \citet{kroup05} highlighted this issue again by examining the anisotropic spatial distribution of the MW dSphs in the context of CDM, pointing out that their distribution can be best represented by a plane highly inclined to the MW stellar disc. \citeauthor{kroup05}\ deemed this to be incompatible with dark-matter dominated satellites, pointing out instead that the plane suggests a causal connection between the satellites. They emphasised that a correlation between the satellites would naturally arise if they were of tidal origin, i.e.\ tidal-dwarf galaxies, originating from an interaction with the young MW.

Tidal-dwarf galaxies \citep[TDG,][]{mirab92} should be found frequently: on the order of ten or more TDGs have been observed to form in single galaxy interaction event in the local Universe \citep[e.g.][]{weilb03,walte06}. The production rate in the early Universe is expected to be even higher given the larger gas fraction in galaxies. \citet{okaza00} argued that if only 1--2 long-lived TDGs form per encounter, the number of dwarf ellipticals (dEs) in very different environments can be explained as being tidal-dwarfs. This number of TDGs per encounter is deduced independently by \citet{delga03} from direct observations strengthening the case for TDGs being a major contributor to the dwarf galaxy population. The formation of TDGs has also been seen in high resolution simulations of gas-rich encounters of galaxies \citep{bourn06,wetzs07}, although very little is known about their fate. It has, however, been shown that dark-matter free dwarf galaxies can survive for several Gyr in a MW like host halo \citep{kroup97}: these computations have shown the existence of long-lived solutions in phase-space that appear dark-matter dominated but are instead derived from DM free satellite galaxies \citep[see also][]{fleck03,metz07b}. The essential point to note is that TDG-formation is an inherent part of \emph{any} cosmological structure formation theory, and sufficient stellar-dynamical and observational work already exists proving their long-term survival. So the natural question to ponder is: where are they in the Local Group?

To constrain the possible origin of dSph satellite galaxies one can study internal processes like radial density- and velocity-dispersion profiles, star-formation histories, or the chemical enrichment of stars. Supplementary to this, their spatial distribution and orbital properties can be investigated. The spatial distribution of the MW satellites has been shown to be highly anisotropic: the satellite galaxies are arranged in a disc-like structure, lying almost perpendicular to the Milky Way's galactic disc \citep{kunke76,lynde76,lynde82,hartw00,kroup05,metz07}. This disc-like alignment, the \emph{disc-of-satellites} (DoS), is statistically very significant and incompatible with being randomly drawn from a distribution of CDM sub-haloes \citep{kroup05, metz07}. Based on the spatial alignment, \citet{kunke76}, \citet{lynde76} and \citet{lynde82} postulated that some of the dSph galaxies, along with several globular clusters (GC), belong to one or two streams of co-moving objects. \citet{lynde83} first suggested that these co-moving dSph galaxies might be tidally torn off a common progenitor galaxy. Incorporating radial velocities, \citet{lynde95} further constrained possible streams of dSphs and GCs and used the great circles of possible locations of angular momenta to identify possible correlations. \citet{palma02} extended the analysis of these so called ``polar-paths'' by using available proper-motion measurements to confine the location of directions of the angular momenta of satellite galaxies to arc-segments of the polar-paths.

Meanwhile further proper motion measurements, especially for more distant dSph galaxies, have become available. In a series of papers \citet{piate03, piate05, piate06, piate07} measured the proper motions of the dSph galaxies Carina, Ursa Minor, Sculptor and Fornax, respectively, with the Hubble Space Telescope (HST). Here we follow \citeauthor{palma02}: we incorporate the \citeauthor{piate03}\ measurements to construct new arc-segments of polar-paths for the MW satellite galaxies in \S\ref{sec_mwsat}. The analysis is concentrated on the 11 ``classical'' (brightest) MW satellites \citep{metz07}, because theoretical (dark matter) modelling of satellite populations has been focused on this sample only. It is also for this sample that velocity information are available. In \S\ref{sec_CDMsim} we analyse, using the techniques developed in \citet{metz07}, the spatial distribution of a set of simulated luminous CDM haloes published with the claim to be a solution to the disc-of-satellite problem. The distribution of the angular momenta of the CDM satellite galaxies are calculated. In \S\ref{sec_pdd} the distribution of the orbital poles for the classical MW and the CDM simulations are statistically analysed. The results are discussed in \S\ref{sec_discuss} and we summarise our conclusions in \S\ref{sec_conclusions}.

\section{Analysis of orbital poles}
\subsection{The Milky Way satellite galaxies}\label{sec_mwsat}
Available proper motion measurements for the satellite galaxies of the MW were collected from the literature, and are given in Table~\ref{tab_mwpropermotion}. We do not give a complete list of all available measurements for the LMC and SMC here, but only the most recent ones. The directions of the angular momenta are not affected by this choice. Multiple measurements of the dSph satellite galaxies are presented where available, however each individual measurement is less precise than the most recent LMC/SMC values. We also indicate whether proper motions were determined using ground based techniques (typically based on old photographic plates with long baselines) or using the Hubble Space Telescope (HST). The published proper motion value for Carina \citep{piate03} was derived without the advanced charge transfer inefficiency (CTI) correction for the STIS camera \citep{brist05}. A re-derived value is given in the table. For the satellites Ursa Minor, Sculptor, and Fornax a weighted mean proper motion using all available measurements was calculated, taking into account all independent single measurements with HST. These mean proper motion values were used in the following analysis.
We note that all measurements for a given satellite are consistent with each other within $3\,\sigma$.

Heliocentric space velocities were calculated from the proper motions and radial velocities from the literature, and corrected for the circular motion of the local standard of rest (LSR) about the Galactic Centre using the IAU recommended value of $V_0 = 220 \kms$ circular velocity \citep{kerr86}, as well as for the peculiar motion of the Sun with respect to the LSR using the velocity components as given by \citet{dehne98}. Positions of the satellite galaxies were transformed to a galactocentric coordinate system \citep[table 1 in][]{metz07}. Having galactocentric positions and velocities the directions of the angular momenta or ``orbital poles'' of the satellite galaxies of the MW were calculated.

The given uncertainties of the proper motion measurements were used to calculate uncertainties of the derived angular momenta. We did not, however, incorporate uncertainties of any other measured quantity: distance and position of satellite galaxies, their radial velocities, the distance of the Sun from the Galactic centre, the circular velocity of the LSR, and the peculiar motion with respect to the LSR. All these uncertainties are typically negligible since they do not contribute much to the total uncertainty of the directions of the angular momenta (see Appendix~\ref{sec_uncert}). Finally, the uncertainties in the direction of the angular momenta were projected onto a unit sphere to derive the arc-segments representing the \emph{projected} $1\sigma$ uncertainties of the directions of the angular momenta \citep[see also][and figure 3 therein]{palma02}.

The orbital poles of the MW's satellites and their corresponding arc-segments are shown in Fig.~\ref{fig_mwam} by the light filled symbols and solid lines, respectively. The dashed loops show regions with $15\degr$ and $30\degr$ distance from the direction of the fitted pole of the DoS for the eleven classical MW satellite galaxies \citep[$l_\oMW = 157\fdeg3,\; b_\oMW = -12\fdeg7$;][]{metz07}. From this figure one can directly identify two satellite galaxies which can not belong to a possible common stream of satellite galaxies: Sagittarius and Sculptor, both having also been excluded by \citet{palma02} as possible stream members. Sagittarius is very close to the Galactic disc and may have precessed sufficiently in the non-spherical potential induced by the Galactic disc, thus being far away from the direction of its initial orbit. Sculptor, however, is more distant and unlikely to be affected much by precession. Compared to the other satellites Sculptor is on a counter-rotating but nearly co-planar orbit about the MW.
The star symbol in Fig.~\ref{fig_mwam} marks the mean spherical direction, $\overline{\vect{l}}$, of the orbital poles of the possible stream members: LMC, SMC, UMi, Dra, Car, and For, $(l_\oMW = 177\fdeg0,\; b_\oMW = -9\fdeg4)$. The small star marks the position of the mean spherical direction if the LMC/SMC centre-of-mass is used instead of both galaxies individually \citep[but see][]{kalli06b}. The solid loop gives the region with a distance corresponding to the spherical standard distance (see \citealp{metz07} for its definition), $\Delta_{\rm sph}=35\fdeg4$, from the mean spherical direction of the sample. \emph{The coincidence of $\overline{\vect{l}}$ and the normal of the DoS provides strong evidence for rotational support.}

The different proper motion measurements from the literature given in Table~\ref{tab_mwpropermotion} sometimes differ by more than $1\sigma$. This is an inevitable weakness of currently available data. Most orbital poles are however not overly affected. Sagittarius is on a perpendicular orbit to the DoS for both $\mu_\delta$ values given in the Table. Likewise, Sculptor is on a counter-rotating orbit for both measurements, ground based and HST, respectively. Only UMi's orbital pole changes when incorporating the independent data, albeit then with very large projected uncertainties. If we take the weighted values for the two HST fields only, its pole-distance is $87.3\degr\mbox{ }^{+31.2\degr}_{-31.0\degr}$, compared to $37.1\degr\mbox{ }^{+8.7\degr}_{-7.7\degr}$ for the mean weighted proper motion of all measurements. Note that the arc-segments are projected uncertainties. While the given proper motion uncertainty for Car is about three times smaller compared to Dra, the projected uncertainty is actually two times larger. Therefore, a large proper motion uncertainty does not necessarily result in a large uncertainty of the location of the orbital pole.

\subsection{CDM simulations}\label{sec_CDMsim}
In the recent search for an explanation of the DoS in terms of CDM substructures, many teams have proposed solutions with varying degrees of means. One of the most recent works claims to be able to arrive at the DoS for all MW-like host haloes \citep{libes05}. For comparison to the Milky Way satellites (\S\ref{sec_mwsat}) this set of simulated systems of luminous satellite haloes in a MW-type dark-matter host halo is analysed. The simulations and galaxy formation models are described in that work and references therein. In short, selected host haloes in a large cosmological dark-matter simulation were re-computed with high-resolution, and galaxy formation was treated by a semi-analytical galaxy formation model (\citealp{cole00}, as extended by \citealp{benso02}). Our data differs slightly from the data used in \citeauthor{libes05} due to an improved semi-analytic galaxy formation model. We analyse a set of the eleven DM sub-haloes containing the most luminous galaxies in each of the six simulated MW-type host haloes. This is a somewhat artificial cut and was chosen by \citet{libes05} to match the number of the ``classical satellite galaxies'' of the MW. The only information used here for the analysis are the final positions and motions of the sub-haloes within the host halo. It is, however, neither sure that the host galaxy is a disc galaxy, nor how its disc would be oriented with respect to the host halo. To be a system comparable to the Milky Way (or Andromeda) both attributes have to be postulated to match with those of the MW.

In a first step the bootstrapping analysis as described in \citet{metz07} was applied to the data, to compare the properties of the spatial distributions. The resulting shape- and strength-parameter are listed in Table~\ref{tab_dhboottest}. None of the simulated sets has as high values as the Milky Way data. Using the same nomenclature as in \citet{libes05}, the simulation gh7 has a slightly higher shape parameter than the MW data, but the strength parameter is significantly lower, which means that the distribution of directions of normals fitted to the bootstrapped data is significantly more widespread than found for the MW. gh2 has a similarly strength parameter but a significantly lower shape parameter; the bootstrapped normals are less circular or show a multi-modal distribution. The other two data-sets with clustered distributions, gh6 and gh10, are lower in both parameters, while gh1 and gh3 have significantly lower shape parameters. Thus none of the simulated MW galaxy haloes appear to reproduce both the shape and strength parameters of the MW's DoS. At a significance level of $1-\alpha=99\%$, all are compatible with the hypothesis to be drawn randomly from a spherical symmetric parent distribution of dark matter sub-haloes.

\begin{figure}
  \plotone{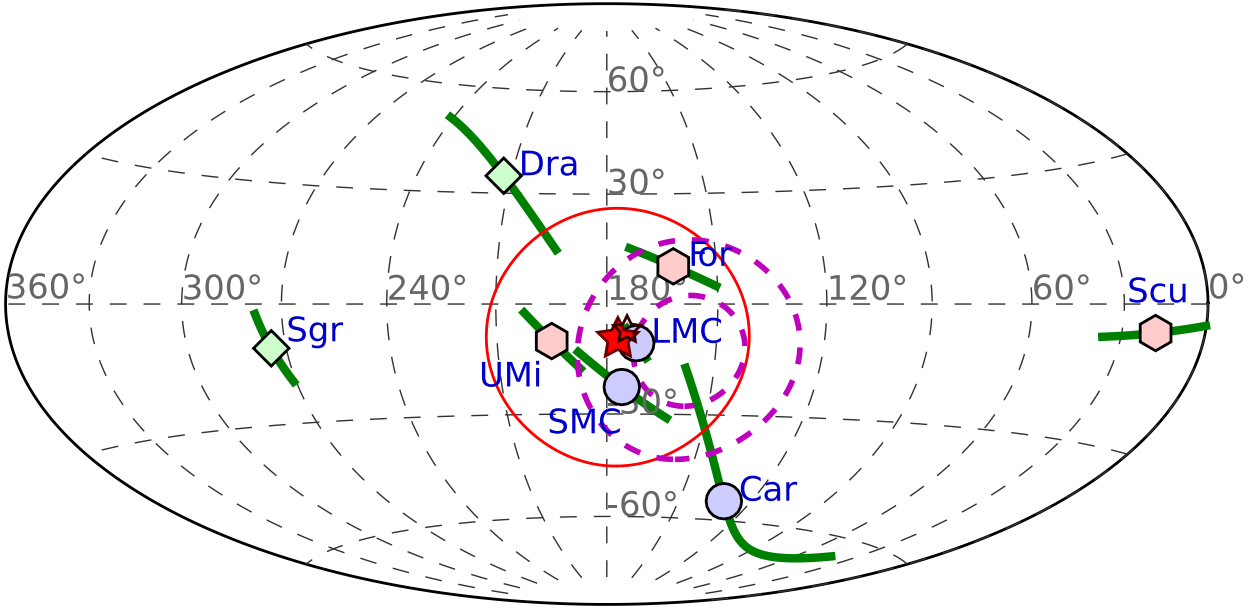}
  \caption{Orbital poles of Milky Way satellite galaxies as derived from their measured proper motion and radial velocities. The directions are shown in an aitoff projection in Galactocentric coordinates. The solid lines give the projected arc-segments derived from the uncertainties of the measured proper motions. Different symbols mark data derived by different methods: circles: HST, diamonds: ground-based measurements, and hexagons: the weighted mean values from Table~\ref{tab_mwpropermotion}. The filled star symbol marks the mean spherical direction $\overline{\vect{l}}$ of the directions of the angular momenta of the satellites excluding Sagittarius and Sculptur, and the solid loop gives the spherical standard deviation of this sample. The smaller, open star symbol marks the mean spherical direction as before, but now treating the LMC/SMC as a bound system whose barycentre is moving with the velocity of the LMC, assuming an LMC/SMC mass ratio of 5/1. 
The dashed loops indicate regions with $15\degr$ and $30\degr$ from the direction of the normal to the plane fitted to the 11 classical Milky Way satellites (the DoS pole). Note the proximity of $\overline{\vect{l}}$ to the normal of the DoS.}
  \label{fig_mwam}
\end{figure}

\begin{figure}
  \resizebox{6cm}{!}{ \plotone{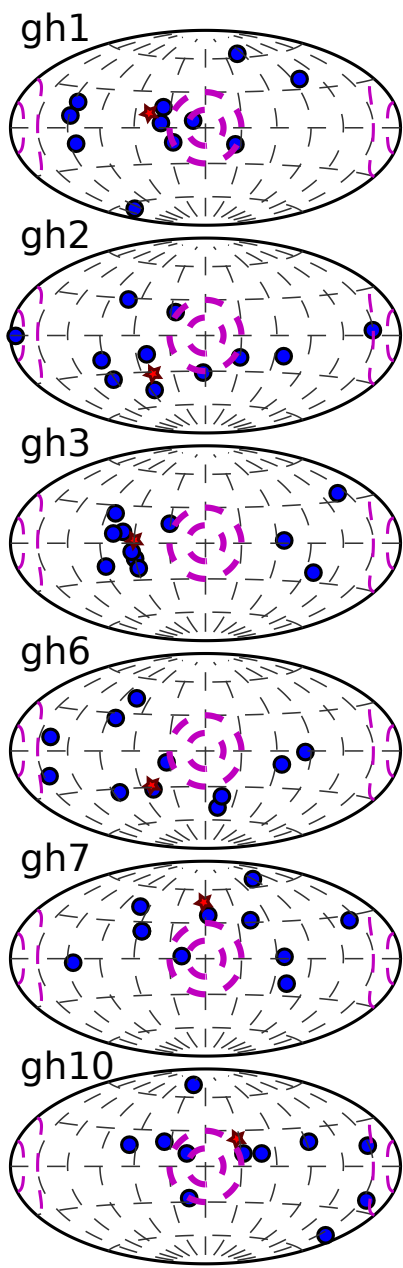} }
  \caption{An aitoff projection of the locations of the orbital poles from the six CDM simulations. The projection is centred on the potential rotational axis -- the normal of the fitted plane -- if the formally fitted disc of satellites were rotationally supported for the simulated data (see text for details). The dashed-loops give the same regions as in Fig.~\ref{fig_mwam}, the star symbol the spherical mean direction of the angular momenta, which is nowhere closer than $\approx 30\degr$ to the normal of the fitted planes.}
  \label{fig_dham}
\end{figure}

To add further insights we compare the distributions of the orbital poles of the satellites. Figure~\ref{fig_dham} shows the directions of the angular momenta of all six simulations. In contrast to Fig.~\ref{fig_mwam} for the MW data, the coordinate system can be arbitrarily chosen, since no galactic disc is comprised in the simulated data which would naturally define the coordinate system. Therefore a coordinate system has been chosen such that the axes of the coordinate system are aligned with the formally fitted plane of the spatial distribution of the simulated satellite galaxies. The projection is chosen such that (i) it is centred on the normal, $\vect{n}$, of the fitted disc - which would be the mean rotational axis \emph{if} the fitted plane were rotationally supported - and (ii) it was chosen such that the plot is centred on the hemisphere that contains the mean spherical direction of the orbital poles $\overline{\vect{l}}$, which is marked by a star symbol. So the plots are centred on the direction
\begin{equation}
\vect{n}_l = \sgn(\vect{n} \cdot \overline{\vect{l}}) \, \vect{n} \label{eqn_nl}
\end{equation}
where $\sgn(x)=x \, |x|^{-1}$.

The choice of the projection for Fig.~\ref{fig_dham} makes the plots also easily comparable with Fig.~\ref{fig_mwam}: If the fitted plane is rotationally supported, all or at least most angular momenta should lie close to the pole of the fitted plane. For comparison the same regions as in Fig.~\ref{fig_mwam} are marked by the dashed loops. A strong clustering in or close to this region is not present in the simulations, and the spherical mean direction, even for selected sub-samples, is nowhere close to the plane's pole.
The agglomeration of five orbital poles seen for the simulation gh1 turned out to belong to a spatially distinct sub-group of sub-haloes at about $200\unit{kpc}$ from the hosts centre that have an approximately common stream motion. This motion may be interpreted as an infalling or bypassing group of sub-haloes that has not yet disbanded \citep[see also][]{li07}. The remaining seven sub-haloes are all found very close to the centre of the host halo in a more or less spherical distribution and the directions of their angular momenta are widely spread.

\subsection{Pole-distance distribution}\label{sec_pdd}
For the MW, as well as for the simulated data, a reference direction $\vect{n}_l$ (Eq.~\ref{eqn_nl}) is given. This would be the approximate orbital pole under the hypothesis that the disc-like structure is rotationally supported, a hypothesis borne out for the MW as we have seen. We construct a pole-distance distribution (PDD) by measuring the angular distances of the orbital poles from this reference direction. \emph{If} the disc is rotationally supported a significant excess of small pole-distances over a random distribution is expected.

\begin{figure}
  \plotone{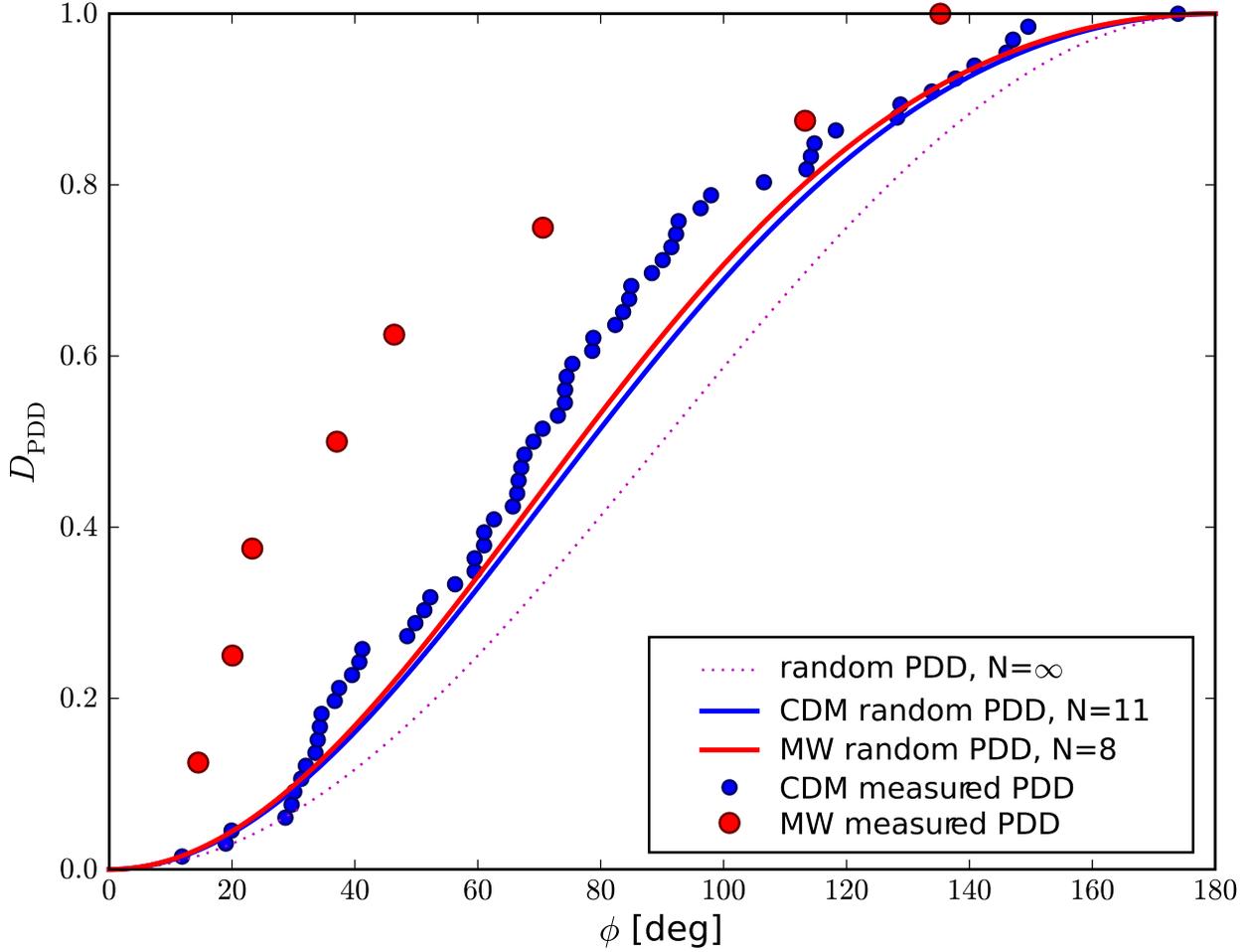}
  \caption{Cumulative distribution function of the pole distance function determined for the Milky Way and the CDM simulated satellite galaxies derived for six MW-halo models from \citet{libes05}. The continuous lines give the theoretical measured PDD for 8 (MW), 11 (CDM), and the dotted line shows the case for an infinite number of random directions (Eqn.~\ref{eqn_PDD}).
}
  \label{fig_poledist}
\end{figure}

In Fig.~\ref{fig_poledist} the cumulative pole-distance distribution for the measured angular momenta of the MW is plotted. For the simulations we build the sum of all individual pole-distance distributions. The cumulative pole-distance function expected for a random distribution, $D_{\rm PDD}(\phi) = \frac{1}{2} ( 1-\cos \phi)$, $0 < \phi < \pi$, is shown by the dotted line. This is, however, not the true parent random distribution for a measured PDD, since the measured PDD is biased: we have chosen that direction of the normal of the fitted plane as the reference pole that is the presumable rotation-pole (\S\ref{sec_mwsat}, \S\ref{sec_CDMsim}). Therefore a biased PDD was constructed: $N$ random directions are created and the mean spherical direction $\overline{\vect{r}}$ of these is determined. As the reference directions the $x$, $y$, and $z$ axes are chosen. The pole-distance is measured from $+\vect{x}$ if the $x$-component of $\overline{\vect{r}}$ is positive, from $-\vect{x}$ otherwise. The same is done for the $y$ and $z$ components. The resulting cumulative biased random PDD is shown in Fig.~\ref{fig_poledist} by solid curves for $N=11$ (CDM) and $N=8$ (MW), respectively. The cumulative biased random pole-distance distribution, created by the Monte Carlo method, can be fitted by
\begin{equation}
  D_{\rm PDD}(\phi, N) = \frac{1}{2} \left( 1-\cos \phi \right) + \frac{0.35}{\sqrt{N}} \sin^2\phi \label{eqn_PDD}
\end{equation}
for $N \ge 5$, $0 < \phi < \pi$. Only in the limiting case $\lim_{N \to \infty}$ Eq.~(\ref{eqn_PDD}) goes over into $D_{\rm PDD}(\phi) = \frac{1}{2} ( 1-\cos \phi)$, which is the purely random, unbiased distribution.

The PDDs were tested against the biased random pole-distance distribution using a Kuiper statistic $V$ and a Watson statistic $U^2$ \citep{steph74}. Both statistics were originally used for distributions on a circle and are thus useable for the PDD. For both statistics and at a significance level $1-\alpha=99\%$ we can not exclude that the PDD for each individual CDM simulated sub-halo system is drawn from a random biased PDD ($N=11$), nor can we exclude that the co-added PDD for all simulated systems is drawn from a random biased PDD. The null-hypothesis can also not be excluded for the Milky Way ($N=8$), but given the very large uncertainties this is not a very strong restriction. Within the uncertainties of the measurements we find the directions of the angular momenta of Dra and Car close to the spherical mean direction ($<\Delta_{\rm sph}$). In that case we can exclude at a very high significance level $1-\alpha=99\%$ that the MW PDD is drawn from a random biased PDD.

Interestingly, the CDM simulated systems show an excess of pole-distances at $\approx 90\degr$, which implies that the fitted disc-like structure tends to disperse. In contrast, we do find an over-density of small pole distances for the MW, suggesting that its DoS is rotationally supported. Note also that we did not choose a reference direction that minimises the pole distances for the Milky Way. If we had chosen the spherical mean direction, the over-density of small pole distances would have been more extreme.

\section{Discussion}\label{sec_discuss}
The bootstrapping analysis of the simulated sub-systems shows them to be compatible with being drawn randomly from a spherical isotropic distribution, whereas this is strongly excluded for the MW satellites \citep{metz07}. Moreover, the analysis of the pole-distance distributions (previous Section) indicates that none of the theoretical CDM sub-structure systems can be rotationally supported, but the orbital poles of all simulated systems are consistent with being drawn from a random distribution. Instead, the claimed ``disc'' of simulated sub-haloes even tends to disperse rather than being supported by rotation. The CDM subhalo ``disc'' is therefore a pressure supported flattened tri-axial spheroid rather than a rotationally supported disc. This finding is supported by \citet{libes07}, who showed that the mean angular momentum of a different set of simulated satellites, using a smooth particle hydrodynamic code, is \emph{not} co-aligned with the normal of a fitted plane to these satellites. Instead it tends to be perpendicular to the long axis of the fitted triaxial ellipsoid, implying that the discs are tumbling around. Note, however, that \citet{libes07} were only able to resolve galaxies brighter than $M_V \approx -12$ which complicates direct comparisons with the MW satellite populations, since most MW dSph are much fainter than that limit.

In contrast, the Milky Way pole-distance distribution, which is still affected by large measurement uncertainties, shows an excess of small pole-distances implying the DoS to be rotationally supported. Within the uncertainties, the two possible DoS satellites with the most deviating directions of the angular momenta, Draco and Carina, are consistent with being found close to the suggested orbital pole of the disc-of-satellites ($<\Delta_{\rm sph}$, Fig.~\ref{fig_mwam}). That case would be highly inconsistent with a random sample ($>99\%$). In order to be able to draw strong conclusions based on the PDD \emph{alone} we need to await more, and more-accurate proper motion measurements. For two satellites, Dra and Car, only a single measurement is available and their projected uncertainties are very large, but measurement campaigns with HST are in progress (Pryor, priv.\ comm.). With upcoming new data, also from the satellite missions Gaia and SIM, the statistics presented here can be used to further test the distribution of orbital poles.

Another fact substantiates the claim of a rotationally supported DoS: the unweighted disc-fitting algorithm \citep{metz07} is most strongly affected by the spatial location of the \emph{outermost} satellite galaxies. We do find, however, a clear alignment of the angular momenta of the \emph{innermost} satellite galaxies with this fitted disc, most strongly for the LMC and SMC. The only exception is Sagittarius whose orbit might already significantly deviate from its initial orbit. It would be rather curious to find a disc-like structure where only the inner satellites orbit within the disc, while the outer ones which mostly define the disc don't. Interestingly, most of the newly discovered dSph also lie close to the DoS (Metz et al., in preparation). It is the combination of these findings, together with the results showing that dSph can be resembled by evolved DM free galaxies \citep{kroup97,metz07b}, that strengthen our notion that the dSph satellite galaxies may not originate from CDM sub-structures.

Our finding for the pole-distance distribution of the Milky Way satellite galaxies poses another challenge to explain them as primordial cosmological substructures, adding to the bunch of already existing issues on galaxy-sized scales. The cusped profiles found in simulations for CDM dominated haloes are in disagreement with the density profiles inferred for MW dSph satellites \citep{gilmo07}, but these are believed to be the most dark-matter dominated objects currently known (\citealp{wilki02,kleyn03}; Dabringhausen, Hilker, \& Kroupa, 2008, submitted). No physical mechanism has been found to be able to evolve cusps to cores \citep{gnedi02}. The missing satellite problem has been addressed by many groups typically explaining the difference in the number of dark-to-light sub-haloes by invoking a suppression of star-formation in low mass sub-haloes. Within the CDM paradigm, the suppression of star formation is usually empirically explained either by feedback processes that heat the halo gas \citep[e.g.][]{kauff93}, or the epoch of reionization which inhibits the formation of small satellites not already seeded \citep[e.g.][]{benso02}. Some recent simulations were able to reasonably reproduce the numbers and luminosities of the bright end of the dwarf galaxies \citep{libes07}. But a complicating factor is that in the literature, depending on which parameters are chosen to match with, different studies use different criteria in the pairing of sub-haloes with satellite galaxies. These are: the most massive sub-haloes \citep{stoeh02}, the most massive progenitors of luminous sub-haloes \citep{libes05}, alternatively the earliest forming haloes \citep{strig07}, low-mass systems that were accreted very recently \citep{sales07a}, or sub-haloes that were grouped before accretion \citep{li07,lake08}. 

Another problem arises for understanding the satellites in terms of DM theory from the models by \citet{besla07}: based on the proper motion measurements for the LMC/SMC \citep{kalli06a,kalli06b} they concluded that if a $\Lambda$CDM-motivated model for the Milky Way halo is assumed, then the LMC/SMC are likely on their first passage about the MW. However, our results show that the orbital poles of both are correlated with the spatial distribution of the dSphs, and also with the orbital poles of UMi, Dra, Car and For, as already suggested by others \citep[e.g.][]{lynde82,lynde95,palma02}. Indeed, the likelihood that an uncorrelated first-time incoming satellite has an orbital angular momentum pointing $9\degr$ away or closer to the mean direction of the orbital poles of the ancient satellites (UMi, For, Dra, and Car, $l_\oMW = 179\fdeg0,\; b_\oMW = -4\fdeg6$) as the LMC does, amounts to 0.6\%. It is therefore very unlikely to find such a strong correlation for completely unrelated satellite galaxies which they would be, if LMC/SMC are on their first fly-by. Either the LMC/SMC proper-motion values are too high, the halo model is incorrect, or an alternative gravitational theory has to be invoked. Among contesters for such theories are MOG \citep{moffa05} and MOND/TeVeS \citep{beken84,beken04}. In modified Newtonian dynamics (MOND) no object can leave the gravitational potential unless it is attracted by a second mass. In this case the LMC/SMC system might be bound to the MW, even though they seem to have a too high velocity. Interestingly, this theory also accounts for the extra dark-matter needed to explain observed rotation curves of TDGs, whereas in Newtonian dynamics an additional mass component is needed \citep{bourn07,genti07}. And, this theory would also account for the observed constant velocity dispersion seen among dSph satellites, $\sigma \approx 10\kms$ \citep{lokas01}, whereby tidal and projection effects would play an important role just as in Newtonian dynamics.

An alternative model to explain the small scale phenomena mentioned above is the tidal-dwarf theory \citep{mirab92,duc94}. The formation of TDGs \citep{zwick56} is observed to commonly occur in gas-rich tidal tails thrown out of interacting galaxies \citep[e.g.][]{hunsb96,duc98,strau06}, and is evident in simulations as well \citep{bourn06,wetzs07}. In contrast to the DoS of the MW, \citet{bourn06} found long-lived TDGs preferentially on prograde, co-planar orbits, but they only considered TDGs more massive than $10^8 \unit{M_\Sun}$, compared to some $10^5$ -- $10^6 \unit{M_\Sun}$ for the least massive dSph satellites of the MW. These models concentrated on low redshift spiral progenitors. Theoretical work has also shown that dark-matter free galaxies can survive for a long time, and can account for most of the observed internal properties (luminosities, morphology, mass-to-light ratios) of dSph satellite galaxies, even in Newtonian dynamics \citep{kroup97, metz07b}. Furthermore, the chemical evolution of TDGs has been studied \citep{recch07}, suggesting that long lasting star-formation episodes can be understood in dark-matter free galaxies. Once TDGs have formed, their chemical evolution is decoupled from each other, which can lead to very different internal evolutions \citep{calur08}. TDGs formed in one encounter, i.e.\ causally-connected TDGs, naturally align within the orbital plane of the interaction, building a rotationally supported disc-like structure. Their distribution may broaden due to precession of orbits and interactions, which also make kick-out scenarios of satellite galaxies due to three-body interaction events likely \citep{zhao98,sales07b}. In this theory, the only exception for the Milky Way satellite galaxies is Sculptor on a counter-rotating orbit, which might result if the tidal tails return to the host phase-shifted. More work is certainly needed to learn about the evolution of (old) TDGs in total: from their birth in a gas-rich early Universe to the possible remnants we see today. The production of TDGs must be an essential part of \emph{any} cosmological theory -- independent of how old TDGs end up: completely destroyed or as observable galaxies.

\section{Conclusions}\label{sec_conclusions}
In this work we have highlighted subtle but convincing arguments that the sub-haloes produced in CDM simulations tailored to address the missing satellite problem can not fully describe the entire population of dSph orbiting the MW. Although our sample is limited to just six simulated galaxy haloes, we nevertheless have demonstrated an 
incompatibility between the spatial distribution of the MW satellites and the most recent CDM simulations that claimed to resemble the MW satellite galaxies. Additionally, the distribution of the orbital poles of the MW satellite galaxies also shows evidence that they differ from the CDM simulated systems. For the simulated haloes the orbital poles of the CDM simulations are consistent with being drawn from a random, isotropic distribution. In contrast, the orbital poles of the Milky Way show an excess of small pole-distances, but are affected by large measurement uncertainties. In fact, the total MW satellite angular momentum is co-aligned with the normal to the DoS-plane implying the DoS to be a rotationally-supported structure -- a feature notably absent in all six simulated haloes. Such a rotationally supported disc is a natural consequence of their formation if the MW dSphs are of tidal origin. Indeed, the production of TDGs is a fundamental process in any hierarchical structure formation theory such that members of this class of objects must be present in the local volume.

Clearly, the problem of the true origin of the MW satellites is of utmost importance for cosmological theory and further tests of models -- CDM and tidal-dwarf -- are required. 
Here we used the currently available proper motion data for the Milky Way satellites, and a limited number of CDM simulated haloes. For further tests of models, and also when new proper motion measurements become available, a fitting function for the biased random cumulative pole-distance distribution is given (Eq.~\ref{eqn_PDD}).

\acknowledgements
We thank an anonymous referee for her/his useful comments that helped to improve the paper. 
We are grateful to Slawomir Piatek for providing us with the re-derived proper-motion values of Carina.

\appendix
\section{On the angular momenta uncertainties}\label{sec_uncert}
To calculate the projected arc-segments of the uncertainties of the direction of the (specific) angular momenta, $\vect{l}=\vect{r} \times \vect{v}$, in Section~\ref{sec_mwsat}, we considered the proper motion uncertainties only, as these contribute to most to the uncertainties. Here we provide the estimation why other uncertainties are negligible. Taking equation (2) from \citet{johns87}, and substituting parallaxes with distances in parsec, we get for the uncertainties of the velocity components in $\mathrm{km\,s^{-1}}$ in Galactic coordinates:
\begin{equation}
\left( \begin{array}{c}
  \sigma^2_U \\ \sigma^2_V \\ \sigma^2_W
\end{array} \right)
 = 
{C} \left( \begin{array}{c}
  \sigma_{v_r}^2 \\
  k^2 \left[ (r \, \sigma_{\mu_{\alpha*}})^2 + (\mu_{\alpha*} \sigma_r)^2 \right] \\
  k^2 \left[ (r \, \sigma_{\mu_{\delta }})^2 + (\mu_{\delta } \sigma_r)^2 \right]
\end{array} \right) + 
  2 \mu_{\alpha*} \mu_{\delta} k^2 \sigma^2_r
\left( \begin{array}{c}
  b_{12} \, b_{13} \\ b_{22} \, b_{23} \\ b_{32} \, b_{33}
\end{array} \right) \; .
  \label{eqn_veluncert}
\end{equation}
$\mu_{\alpha*}$, $\mu_{\delta}$ are the proper motion components in right ascension and declination, respectively, given in $\mathrm{arcsec\;yr^{-1}}$, $r$ is the distance in $\mathrm {pc}$, and $v_r$ is the radial velocity in $\mathrm{km\,s^{-1}}$. The constant $k=4.74057$, and the elements $b_{ij}$ are the components of the matrix $B$ as given by \citet{johns87}. $B$ is a rotation matrix, performing the transformation from radial velocities and proper motions to velocity components in Galactic coordinates. $C$ is a matrix where each element is the square of the individual elements of the matrix $B$: $c_{ij}=b_{ij}^2$. 

The second and third component in the first term of the velocity uncertainty vector in Eqn.~(\ref{eqn_veluncert}) are given by: $\left( k^2 \left[ (r \sigma_\mu)^2 + (\mu \sigma_r)^2 \right] \right)$. With $\sigma_\mu \sim 0.1 \mu$ and $r \gg \sigma_r$, typical values are $\sigma_r \sim 0.01 r$, we get $k^2 \mu^2 (0.01 r^2 + 0.0001 r^2) \approx k^2 \sigma_\mu^2 r^2$, which results in a velocity uncertainty of the order of $\sim 25\kms$ for each component at a distance of $50\unit{kpc}$. For the second term, with $\mu_{\alpha*} \approx \mu_\delta$, it follows $2 k^2 \mu^2 \sigma_r^2 \ll k^2 \sigma_\mu^2 r^2$. $B$ is a rotation matrix, with $b_{ij} \, b_{ik} \le 1$, so it is reasonable to neglect the distance uncertainties.

Radial velocity uncertainties are typically of the order of $5\kms$, compared to $\sim 25\kms$ for the two transverse velocity components as estimated above. The absolute value of the velocity uncertainty is then estimated to be of the order $2 \cdot (25\kms)^2 + (5 \kms)^2 \approx 2 \cdot (25\kms)^2$. Moreover, as we calculate the specific angular momentum vector given by $\vect{l} = \vect{r} \times \vect{v}$, the contribution to the vector $\vect{v}$ from the radial velocity is almost parallel to the vector $\vect{r}$. The angle is at maximum of the order $\arctan(8.5\unit{kpc} / 50\unit{kpc}) \approx 10\degr$, so the radial velocity uncertainty contributes only little to the total velocity uncertainty.

For the distance to the Galactic Centre, we use the IAU recommended value $8.5\unit{kpc}$, while some authors prefer a value of $8\unit{kpc}$, so we assume the uncertainty to be $0.5\unit{kpc}$. This uncertainty influences only the first component of the position vector $\vect{r}$ in Galacocentric coordinates. The variance of the second component of the specific angular momentum vector is given by
$\sigma_{l_2}^2 = \left(r_3^2 \sigma_{v_1}^2 + v_1^2 \sigma_{r_3}^2\right) + \left(r_1^2 \sigma_{v_3}^2 + v_3^2 \sigma_{r_1}^2\right)$. The contribution due to the velocity uncertainties is of the order $2 \cdot \left(30 \unit{kpc}\right)^2 \left(25\kms \right)^2 = 1\,125\,000 \left( \unit{kpc} \kms \right)^2$, while the distance uncertainty to the Galactic Centre contributes with only about one thousandth of this: $\left(100 \kms \right)^2 \left(0.5 \unit{kpc} \right)^2 = 2\,500 \left( \unit{kpc} \kms \right)^2$. Furthermore, the distance uncertainties to a satellite at $50\unit{kpc}$ is of the order $2\unit{kpc}$, so for each component of the order of $\sqrt{2^2/3}\unit{kpc} \approx 1.15 \unit{kpc}$. These do also not contribute significantly to the total uncertainty.
Finally, the transverse position uncertainties in $\alpha$ and $\delta$ are much smaller than the distance uncertainties. At a typical distance of a satellite galaxy of the order of $50\unit{kpc}$, a position uncertainty of $1\arcmin$ transforms into a spatial uncertainty of $\approx 15\unit{pc}$, which is two orders of magnitudes smaller than a typical distance uncertainty of $2\unit{kpc}$.

For the circular motion of the LSR about the Galactic Centre we use the IAU recommended value of $220\kms$. The uncertainty of this value is of the order $20\kms$, so this might have a significant influence on the location of pole of the angular momentum vector. We checked that the results do not change significantly by once using $200\kms$ and once $240\kms$ for the circular motion. The uncertainties of the peculiar motion of the Sun with respect to the LSR as given by \citet{dehne98} are of the order of $0.5\kms$, much smaller than the other contributions and can be neglected.

Sagittarius is the closest satellite considered here. With a distance of $24\unit{kpc}$ from the Sun, the estimations given above might not be valid anymore. But its geometric location is very special, as it is located almost directly behind the Galactic Centre, and $\vect{r} \| \vect{v}_r$. Moreover, Sgr is anyway not co-rotating with the DoS because of its spatial location with respect to the DoS.

\bibliographystyle{apj}
\bibliography{quotes,quotebooks,quotes3,quotes2,quotes_preprint,quotesobservations}

\clearpage
\begin{table*}
\caption{Measured proper motions of the satellite galaxies of the Milky Way used for the analysis. In the second column the galacto-centric distance of the satellite galaxies to the MW is given. The fifth column indicates whether values were measured using ground based telescopes (GND) or the HST. Weighted mean values (MEAN) are indicated as well.}\label{tab_mwpropermotion}
{\centering
  \begin{tabular}{lccccl}
  \tableline
  Name & $d_{\rm GC}$ & $\mu_{\alpha} \cos \delta$ & $\mu_\delta$ & & Reference\\
       & (kpc) & $\left(\unit{mas\; yr^{-1}} \right)$ &
         $\left(\unit{mas\; yr^{-1}} \right)$ &
         \\
  \tableline
  Sgr &  16 & $-2.65 \pm 0.08$   & $-0.88 \pm 0.08$  & GND & \citet{ibata97}\\
      &     & $-2.83 \pm 0.20$   & $-1.33 \pm 0.20$  & GND & \citet{dines05}\\
  LMC &  50 & $+2.03 \pm 0.08$   & $+0.44 \pm 0.05$  & HST & \citet{kalli06b}\\
  SMC &  57 & $+1.16 \pm 0.18$   & $-1.17 \pm 0.18$  & HST & \citet{kalli06a}\\
  UMi &  68 & $+0.5 \pm 0.8$     & $+1.2 \pm 0.5$    & GND & \citet{schol94}\\
      &     & $+0.056 \pm 0.078$ & $+0.074 \pm 0.099$& GND & \citet{schwe97}\\
      &     & $-0.50 \pm 0.17$   & $+0.22 \pm 0.16$  & HST & \citet{piate05}\\
      &     & $-0.04 \pm 0.07$   & $0.13 \pm 0.07$   & MEAN & \\
  Scu &  79 & $+0.72 \pm 0.22$   & $-0.06 \pm 0.25$  & GND & \citet{schwe95}\\
      &     & $+0.09 \pm 0.13$   & $+0.02 \pm 0.13$  & HST & \citet{piate06}\\
      &     & $+0.25 \pm 0.11$   & $ 0.00 \pm 0.12$  & MEAN & \\
  Dra &  82 & $+0.6 \pm 0.4$     & $+1.1 \pm 0.5$    & GND & \citet{schol94}\\
  Car & 103 & ($+0.22 \pm 0.09$   & $+0.15 \pm 0.09$)& HST & \citet{piate03}\tablenotemark{a}\\
      &     & $+0.22 \pm 0.13$   & $+0.24 \pm 0.11$  & HST & Piatek (2007, priv. comm)\\
  For & 140 & $+0.59 \pm 0.16$   & $-0.15 \pm 0.16$  & GND & \citet{dines04} \\
      &     & $+0.476 \pm 0.046$ & $-0.36 \pm 0.041$ & HST & \citet{piate07} \\
      &     & $+0.485 \pm 0.044$ & $-0.354 \pm 0.028$ & MEAN & \\
  \tableline
  \end{tabular}
} 
\tablenotetext{a}{Proper motion measurement without advanced CTI correction for HST/STIS.}
\end{table*}

\clearpage
\begin{table}
\caption{Shape parameter $\gamma$ and strength parameter $\zeta$ of the distribution of directions of normal vectors fitted to 10\,000 bootstrap samples of the simulated CDM satellite galaxies. The first column gives an identifier of the simulated haloes. For comparison, the values for the Milky Way are given in the last row. See \citet{metz07} for further details.}
\label{tab_dhboottest}
{\centering
  \begin{tabular}{lcc}
  \tableline
   & shape parameter $\gamma$ & strength parameter $\zeta$ \\
  \tableline
  gh1  & 0.5 & 3.2 \\
  gh2  & 2.3 & 4.2 \\
  gh3  & 0.3 & 3.5 \\
  gh6  & 1.2 & 3.0 \\
  gh7  & 3.9 & 2.6 \\
  gh10 & 1.5 & 2.8 \\
  \tableline
  MW   & 3.85 & 4.29 \\
  \tableline
  \end{tabular}
}
\end{table}

\clearpage

\clearpage

\clearpage

\end{document}